\documentclass[12pt, a4paper]{article}
\usepackage{graphicx}
\usepackage{color}
\newtheorem{lemma}{Lemma}

\title{\bf Slow passage through parametric resonance for a weakly nonlinear dispersive
wave\thanks{ This work was supported by grants RFBR 03-01-00716 and DFG TA 289/4-1.}}

\author{Sergei Glebov\thanks{Ufa State Petroleum Technical University({\tt sg@anrb.ru})}
\and Oleg Kiselev \thanks{Institute of Mathematics, USC RAS ({\tt ok@ufanet.ru})}
\and Nikolai Tarkhanov \thanks{Institute of Mathematics, Potsdam University({\tt tarkhanov@math.uni-potsdam.de}).}}
\date{}
\begin{document}
\maketitle

\begin{abstract}
A solution of the nonlinear Klein-Gordon equation perturbed by a parametric driver is studied.
The frequency of the parametric perturbation varies slowly and passes through a resonant value. 
It yields a change in a solution. We obtain a connection formula for the asymptotic 
solution before and after the resonance. 
\end{abstract}
%\begin{keywords}
%nonlinear optics, resonance, solitons, nonlinear waves\end{keywords}
%\begin{AMS}
%35Q60, 37K40, 78M35\end{AMS}

\font\Sets=msbm10
\def\Real{\hbox{\Sets R}}
\def\Complex{\hbox{\Sets C}}
\def\bb{\begin{equation}}
\def\ee{\end{equation}}
\def\pt{\partial}
\def\mod{\hbox{mod}}
\def\const{\hbox{const}}
\def\sgn{\hbox{sgn}}
\def\Arg{\hbox{Arg}}
\def\a{\alpha}
\def\b{\beta}
\def\d{\delta}
\def\G{\Gamma}
\def\g{\gamma}
\def\e{\epsilon}
\def\ve{\varepsilon}
\def\k{\kappa}
\def\l{\lambda}
\def\O{\Omega}
\def\o{\omega}
\def\th{t}
\def\s{\sigma}
\def\t{\tau}
\def\z{\zeta}
\newtheorem{theorem}{\bf Theorem}
\def\qed{\vrule height 7pt width 7pt depth 0pt}

\section{Introduction}
This work is devoted to the problem on a control of a nearly monochromatic weakly nonlinear dispersive wave with small amplitude in a strong nonlinear media.  It is well-known that packets of nearly monochromatic waves propagate without changing of their shape when the envelope function of the packet is a soliton of the Nonlinear Schr\'odinger equation(NLSE). The solitary packets of waves would be more suitable for communication in optical fibers on a large distance if one can control the parameters of the envelope function for such packets. The wave packets with a soliton-like shape form for sufficiently large range of initial data. The control of the shape is possible by a parametric perturbation of the system. 
\par
Here we propose a new approach for controlling of parameters of the solitary packets. In our approach the wave packets are controlled due to a slowly passage through a local parametric resonance. 
\par
In this paper we study the parametrically driven  nonlinear Klein-Gordon equation. The frequency of the driver varies slowly. It is well-known that an envelope function of weak nonlinear wave is a solution of the NLSE  \cite{kelley}-\cite{zaharov}. In general case this is valid proposition for the parametrically driven Klein-Gordon equation. But the small driving force plays a central role in narrow layers where the frequency passes through the resonant value. As a result fast and slow variables appear which define the behaviors of the solution. In these layers the envelope function of a primary parametric resonance equation instead of the NLSE. In the slow variables the passage through the resonant layer looks as a jump of the envelope function. Our goal is a connection formula for this jump.
\par
The change of the solution in the resonant layers for a local resonance (not parametric) was studied in \cite{kevorkian}-\cite{sg}. The detail description of the phenomenon was presented in \cite{kg1}. In that case the solution in the resonant layer is defined by the Fresnel integral. For parametric resonance the solution was studied in \cite{BuslaevDmitrieva} for the Mathieu  equation. They shown that the solution defines by parabolic cylinder functions. In our work we found that the parabolic cylinder equation also allows one to obtain the connection formulas for the small solutions of nonlinear Klein-Gordon equation.
\par
In this work we use the singular perturbation theory and matching of different asymptotic expansions to obtain the connection formula. Our approach is based on the matching method of asymptotic expansions \cite{Ilin}.
\par
The structure of the paper is as follows. In Section 2 we formulate the result. The numerical simulations are inserted in Section 3.  Section 4 contains the formal construction for the asymptotic solution out of the resonant layers. In Section 5 the asymptotic solution is constructed in the resonant layer. In Section 6 we  match the main terms of the  asymptotic expansions.

\section{Statement of the problem and main result}

We study  the Klein-Gordon equation with a cubic nonlinearity 
\begin{equation}
\partial^2_{t} U - \partial^2_{x} U + \bigg(1+\ve f\cos\bigg(\frac{S(\ve^2x,\ve^2t)}{\ve^2}\bigg)\bigg)U + \gamma U^3 =0,
\quad 0 < \ve \ll 1. \label{pdkg} 
\end{equation}
Here $\gamma$ and $f$ are constants.  The phase function $S(y,z)$ and all derivatives of $S(y,z)$ are bounded. 
\par
Our goal is to obtain an asymptotic solution for (\ref{pdkg}). To formulate the result we use the following notations for slow variables:
$$
x_j = \ve^j x,\quad t_j = \ve^j t, \quad j=1,2.
$$
Let us define
$$
L[\chi]=\big[\pt_{t_2}\chi\big]^2-\big[\pt_{x_2}\chi\big]^2-1
$$
and
$$
l(x_2,t_2)\equiv L[\omega t_2+kx_2+S].
$$
\par
The main result of the paper is formulated in the Theorem.
\begin{theorem}\label{mainTheorem}
In the domain $l<0$ the formal asymptotic  solution of (\ref{pdkg}) has the form
$$
U(x,t,\ve) \sim \ve u_1(x_1,t_1,t_2)\exp\{i(kx+\omega t) \}+c.c..
$$
The amplitude $u_1=\Psi \exp\left\{-i\displaystyle\frac{f^2}{4\omega}G(x_2,t_2)\right\},$ where $G$ is an antiderivative of 
$$ 
g(x_2,t_2)=\left[\frac{1}{L[\omega t_2+kx_2-S]} + \frac{1}{L[\omega t_2+kx_2+S]}\right], 
$$
with respect to $t_2$.
\par
Function $\Psi$ is determined by the nonlinear Schr\"odinger equation 
$$
i\omega \pt_{t_2}\Psi - \pt_{\zeta}^2\Psi +3\gamma |\Psi|^2 \Psi = 0,\quad \zeta=\omega x_1+kt_1.
$$
In the domain $l>0$ the formal asymptotic solution has the same form
$$
U(x,t,\ve) \sim \ve v_1(x_1,t_1,t_2)\exp\{i(kx+\omega t) \}+c.c..
$$
The amplitude $v_1=\Psi  \exp\left\{i\displaystyle\frac{f^2}{4\omega}G(x_2,t_2)\right\}$ and $\Psi$ is determined by the nonlinear Schr\"odinger equation also  and initial datum on the curve $l=0$
$$
\Psi(t_2,\zeta)|_{l=0}=e^{{f^2\pi\over8}}\psi +
{e^{i\pi/4}e^{{f^2\pi\over16}}
e^{-i{f^2\over8}\ln(2)}f\sqrt{\pi}
\over 2\Gamma(1-i{f^2\over8})} 
\overline{\psi},
$$
where $\psi=\Psi(t_2,\zeta)|_{l=-0}$
\end{theorem}

\section{Numerical Simulation}

\par
Here we illustrate the analytical results which are formulated in Theorem \ref{mainTheorem}. Let us consider  (\ref{pdkg}) with $\gamma=-1/3$ and $S(t_2)=t_2^2/2$. We consider the solution when $k=0$. It yields $\omega=1$ and $\zeta=x_1$. The envelope function is a solution of the nonlinear Schr\"odinger equation:
$$
2i\pt_{t_2}\Psi-\pt_{x_1}^2\Psi-|\Psi|^2\Psi=0.
$$ 
It is well-known that this equation has a soliton solution:
$$
\Psi(x_1,t_2)={\sqrt{2}\eta e^{i(\kappa x_1+(\kappa^2-\eta^2)t_2/2)}\over \cosh(\eta(x_1+ \kappa t_2))}.
$$
\par
In this section we reduce the problem and take into account the envelope function as $\eta=1$ and $\kappa=0$. In this case the asymptotic solution before resonance has the following form:
$$
u\sim \varepsilon{\sqrt{2}e^{i t_2/2}\over \cosh(x_1)}e^{i\big(kx+\omega t-{f^2\over8}\ln\big({t_2-2\over t_2+2}\big)\big)}.
$$
\par
In this simplest case the resonant curves are $t_2=-2$, $t_2=0$ and $t_2=2$. Let us consider $\varepsilon=0.1$ then the resonant curves are $t=-200$, $t=0$ and $t=200$. Following figures show the behavior of the solution for (\ref{pdkg}). Initial data are represented as: 
\begin{eqnarray*}
u|_{t=-900}=\varepsilon\bigg({\sqrt{2}e^{i \varepsilon^2 t/2}\over \cosh(\varepsilon x)}e^{it-{f^2\over8}\ln\big({t\varepsilon^2-2\over t\varepsilon^2+2}\big)\big)}\bigg)\bigg|_{t=-900},\\
\partial_t u|_{t=-900}=\varepsilon\partial_t\bigg({\sqrt{2}e^{i \varepsilon^2 t/2}\over \cosh(\varepsilon x)}e^{i\big(t-{f^2\over8}\ln\big({t\varepsilon^2-2\over t\varepsilon^2+2}\big)\big)}\bigg)\bigg|_{t=-900}.
\end{eqnarray*}
\par
\begin{tabular}{c}
\includegraphics[width=13cm,height=13cm]{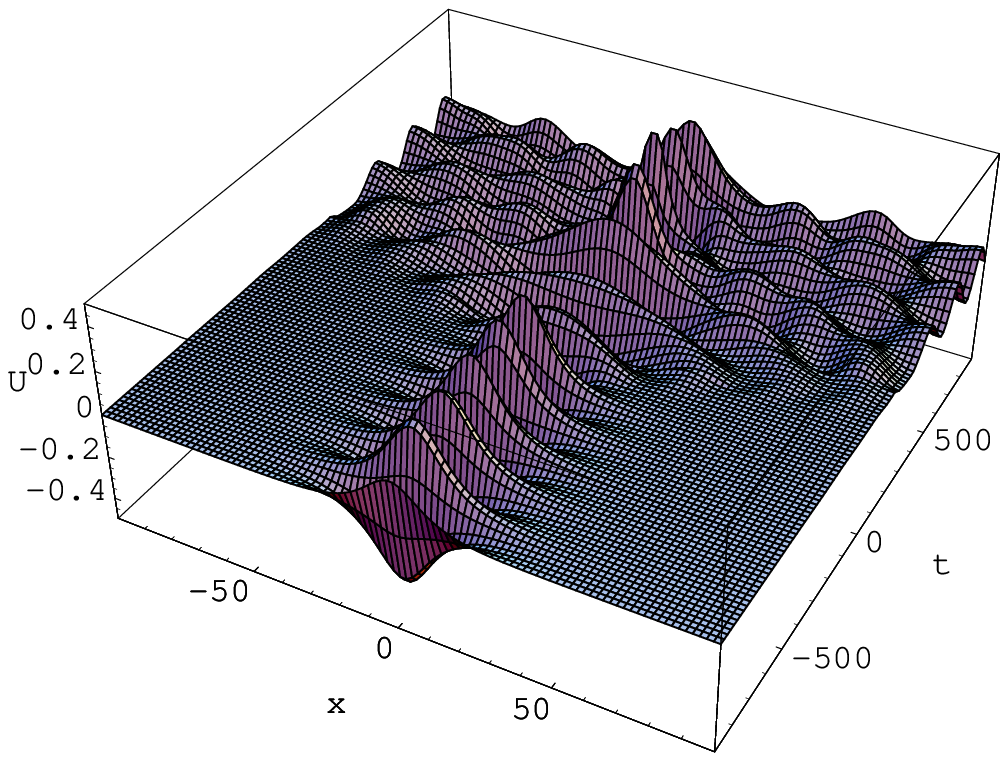} \\
{Fig.1. 3-Dimensional behaviour of numerical solution of (\ref{pdkg})}
\end{tabular}
\par
\begin{tabular}{c}
\includegraphics[width=13cm,height=13cm]{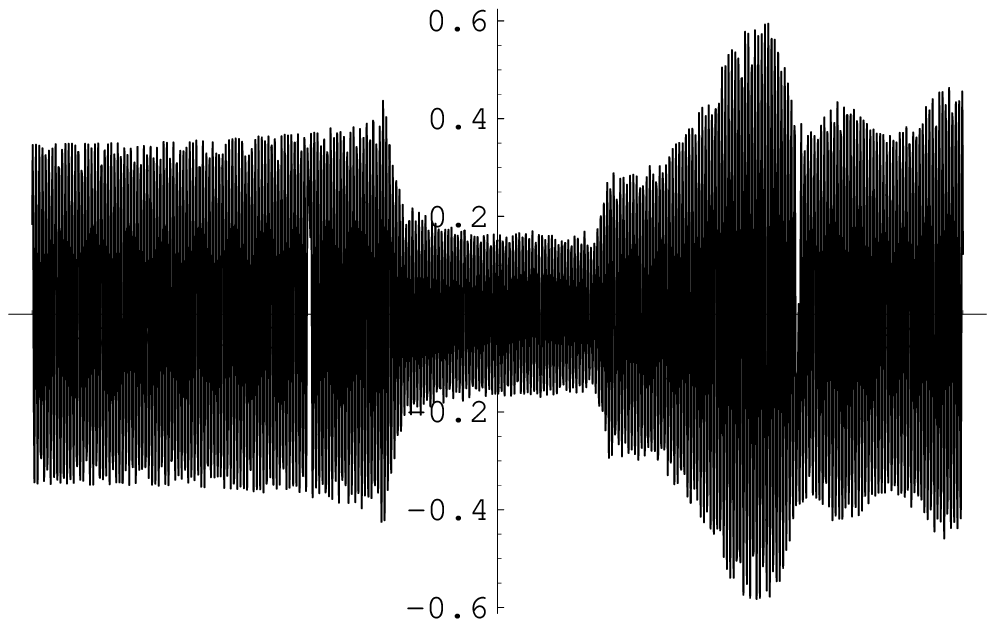}\\
{Fig.2. Profile of the numerical solution when $x=0$}
\end{tabular}
\par

\section{The first external expansion}
\label{externalAsymptotics1}
\par
In this section the formal 
asymptotic solution is constructed 
in the domain before the resonance. 
This domain is defined by the 
condition $l<0$. The asymptotic 
expansion has the form of WKB-type. 
The leading-order term of asymptotic 
expansion has the order 
of $\varepsilon$  and  oscillates.
\par
\begin{theorem}\label{ex1Theorem}
When $l<0$ the formal asymptotic 
expansion for the  solution of 
(\ref{pdkg})  modulo $O(\ve^N)$ 
has the form
\begin{equation}
U=\sum_{n\ge 1}^N \ve^n {U}_n,
\quad N\in{\mathbf N}\label{ext1}
\end{equation}
where
\begin{equation}
U_1 = \sum_{j=\pm1}u_{1(j,0)}(t_1,x_1,t_2)
\exp\{i j(kx_2+\omega t_2)/\varepsilon^2\}
\label{firstCorrection}
\end{equation}
and
$$
U_n=\sum_{(j,m)\in \Omega_n}u_{n,(j,m)}(t_1,x_1,t_2)\exp(i\chi_{j,m}(x_2,t_2)/\ve^2). 
$$
Here $\chi_{j,m}=j(kx_2+\omega t_2)+mS$ 
is the phase of the ocsillating
mode and the set $\Omega_n$ 
contains the pairs $(j,m)$ such that:
\begin{eqnarray*}
\Omega_1=&
\{(\pm1,0)\},
\\
\Omega_n=&
\{(j_{n-1},m_{n-1}\pm1),(j_{n-1},m_{n-1})\in\Omega_{n-1}\}\cup
\\
&
\{(j_{l_1}+j_{l_2}+j_{l_3},m_{l_1}+m_{l_2}+m_{l_3}),
\\
&
l_1+l_2+l_3=n,\,(j_{l_q},m_{l_q})\in\Omega_{l_q},l_q\in{\mathbf N}\}.
\end{eqnarray*}
The coefficient $u_{1(1,0)}$ is defined 
out of (\ref{nlse11}). The 
higher-order terms $u_{n,(j,m)}$ are defined 
by linearized equation (\ref{n-th-1-n-2}) 
as $(j,m)=(\pm1,0)$ and by algebraic 
equations (\ref{n-th-j-m}) as 
$(j,m)\not=(\pm1,0)$.
\par
This expansion is valid in the 
domains
$$
-\varepsilon^{-1}L[\chi_{j,m}] \gg 1, 
\quad \forall(j,m)\in\cup_{q=1}^{N}\Omega_q.
$$
\end{theorem}
\par
This theorem give the asymptotic solution for a set of the domains on the plane $(x_2,t_2)$. But this theorem does not give an answer about a connection between the asymptotic solutions for nearest-neighbour domains.
\par
This theorem does not give the solution in the form  asymptotic series for (\ref{pdkg}) in general. Because the resonant lines $L[\chi_{j,m}]=0$  are dense as $(j,m)\in {\mathbf Z^2}$ and full series (\ref{ext1}) does not asymptotical anywere.
\par
Let us consider an example as  $S\equiv t_2^2/2$ and $k=0$. The set $\Omega_n$ contains a phases $\chi_{j,m}$ as $j=2q+1$ and $q,m=\pm0,1,2,\dots$, $|j|+|m|<n$. In this case $\omega=1$ 
and the lines 
$$
1-(2q+1)^2+m[-2(2q+1)t_2-mt_2^2]=0
$$
are resonant. These lines
$$
t_2=-{2q+1\over m}\pm {1\over m}
$$
are dance everywhere when $N=\infty$. But for any segments of (\ref{ext1}) 
the sets of resonant lines are finite. When $N=3$ the set is $t_2=-2$,$t_2=0$ and $t_2=2$.

\subsection{Construction of pre-resonant solution} 
\label{subsectionPre-ResonantAsymptotics}
\subsubsection{Derivation of equations}
Let us substitute (\ref{ext1})  in equation (\ref{pdkg}) and gather the terms of same order of $\ve$. As a result we obtain a recurrent sequence of the equations:
\begin{eqnarray}
\partial_t^2U_n-\partial_x^2U_n +U_n =
& -f\cos\big({S\over\varepsilon^2}\big)
U_{n-1} -\nonumber
\\
&\gamma\sum_{j+l+m=n}U_jU_lU_m 
-2\partial_{t_1}\partial_tU_{n-1}
\nonumber\\
&+
 2\partial_{x_1}\partial_xU_{n-1}
- \partial_{t_1}^2U_{n-2}+ 
\partial_{x_1}^2U_{n-2}
\nonumber\\
&-2\partial_{t_2}\partial_tU_{n-2}
-2\partial_{t_1}\partial_{t_2}U_{n-3} -
\partial_{t_2}^2U_{n-4}.
\label{n-thPreresonantEq}
\end{eqnarray}
\par
Let us consider this equation at $n=1$:
$$
\bigg(-\omega^2+k^2+1\bigg)u_{1,(1,0)}
\exp(i(kx+\omega t))+c.c.=0.
$$
This equation defines the 
dispersion formula:
$$
\omega^2=k^2+1.
$$
Equation at $n=2$ is 
\begin{eqnarray}
\sum_{(j,m)\in \Omega_2}
&
\bigg[\bigg(-(\pt_{t_2}\chi_{j,m})^2+
(\pt_{x_2}\chi_{j,m})^2+1\bigg) 
u_{2,(j,m)}\exp(i\chi_{j,m}/
\ve^2)\bigg]=
\nonumber \\
&
-\bigg(2i\omega 
\pt_{t_1}{u}_{1,(1,0)}-
2ik\pt_{x_1}{u}_{1,(1,0)} 
\bigg)\exp(i(kx+\omega t))-  
\nonumber \\
&
f\cos(S/\ve^2+\phi_0)u_{1,(1,0)}
\exp(i(kx+\omega t))+ c.c.,
\label{eqAsEps2}
\end{eqnarray}
where $\Omega_2={(j,m),\,j=\pm1,\,m=\pm1}$.
\par
We construct a bounded solution respect to fast time $t$. We remove terms with modes $\exp(i(kx+\omega t))$ from the right-hand side of (\ref{eqAsEps2}). It yields:
\begin{equation}
\omega \partial_{t_1}{u}_{1,(1,0)}-
k\partial_{x_1}{u}_{1,(1,0)}=0.\label{u1Eq}
\end{equation}
Solution of this equation is an arbitrary function with respect to $\zeta=kt_1+\omega x_1$.
\par
The set of of $\Omega_2$ is defined by modes of right-hand side of (\ref{eqAsEps2}). 
The amplitudes for these modes are given by following formulas:
\begin{equation}
\bigg(-(\pt_{t_2}\chi_{j,m})^2+
(\pt_{x_2}\chi_{j,m})^2+1\bigg)u_{2,(j,m)}
=-\frac{1}{2} f u_{1,(1,0)}, 
\label{algE2}
\end{equation}
where $(j,m)\in\{(1,\pm1)\}$ and 
complex conjugated equations
for  $(j,m)\in\{(-1,\pm1)\}$. 
Thus we define $U_2$. 
\par
It is easy to see that  $U_2$ has singularity of first order on the curves:
\begin{eqnarray}
L[(k x_2+\omega t_2)+S]=0
\label{singCurvePlus1}
\\ 
L[(k x_2+\omega t_2)-S]=0.
\label{singCurveMinus1}
\end{eqnarray}
\par
Equation as $\ve^3$ looks like
\begin{eqnarray}
\pt_t^2 U_3-\pt_x^2 U_3+U_3=
+2i\omega \pt_{t_2}{u}_{1,(1,0)}
\exp(i(kx+\omega t))
\nonumber\\
 +
\sum_{(j,m)\in \Omega_2}\bigg[
2i\pt_{t_2}\chi_{j,m} \pt_{t_1}u_{2,(j,m)}-
\nonumber\\
2i\pt_{x_2}\chi_{j,m} \pt_{x_1}u_{2,(j,m)}
\bigg]\exp(i\chi_{j,m}/\ve^2) +
\nonumber\\
(\pt_{t_1}^2u_{1,(1,0)}-\pt_{x_1}^2u_{1,(1,0)})
\exp(i(kx+\omega t))+
\nonumber\\
f\cos(S/\ve^2)\sum_{(j,m)\in \Omega_2}
u_{2,(j,m)}\exp(i\chi_{j,m}/\ve^2)
\nonumber\\
+ \gamma \bigg[u_{1,(1,0)}^3
\exp(3i(kx+\omega t))+
\nonumber\\
3|u_{1,(1,0)}|^2u_{1,(1,0)}
\exp(i(kx+\omega t))\bigg]+
c.c.=0.
\label{eqAsEps3}
\end{eqnarray}
This equation has a bounded solution with respect to $t$ if the right-hand side does not contain the terms with $\exp(\pm i(kx+\omega t))$. Thus we obtain two equations. First of them defines the dependence of $u_{2,(\pm1,0)}$ 
with respect to $x_1$ and $t_1$:
\begin{equation}
\omega \partial_{t_1}{u}_{2,(1,0)}-
k\partial_{x_1}{u}_{2,(1,0)}=0.\label{u1Eq}
\end{equation}
The second equation defines the dependence of  $u_{1,(1,0)}$ with respect to $t_2$ and $\zeta$:
\begin{eqnarray*}
2i\omega \partial_{t_2}{u}_{1,(1,0)} - 
\partial_{\zeta}^2u_{1,(1,0)} +
\\
3\gamma |u_{1,(1,0)}|^2 u_{1,(1,0)} =  
\frac{f^2}{4}g(x_2,t_2) u_{1,(1,0)}, 
\label{nlse11}
\end{eqnarray*}
where 
$$
g(x_2,t_2) =
\left[\frac{1}{L[\omega t_2+kx_2-S]} + \frac{1}{L[\omega t_2+kx_2+S]}\right]
$$
\par
Let us denote 
\begin{equation}
u_{1(1,1)}=\Psi\exp\left\{-i\frac{f^2}{8\omega}G(x_2,t_2)\right\},
\label{formulaG}
\end{equation}
where $G$ is an antiderivative of $g(x_2,t_2)$  with respect to $t_2$. Then the function $\Psi$ is determined by 
the nonlinear Schr\"odinger equation 
$$
2i\omega \pt_{t_2}\Psi - \pt_{\zeta}^2\Psi +3\gamma |\Psi|^2 \Psi = 0.
$$
\par
The coefficients with $\exp(i(k_2x+\omega t_2\pm S)/\varepsilon^2)$ 
and their complex conjugations are defined by:
\begin{eqnarray}
 u_{3,(j,\pm1)}={1\over L[\chi_{j,\pm1}]} 
\bigg[2i\pt_{t_2}\chi_{j,\pm1} 
\pt_{t_1}u_{2,(j,\pm1)}-
2i\pt_{x_2}\chi
\pt_{x_1}u_{2,(j,\pm1)}\bigg],\quad j=\pm1.
\label{algU31}
\end{eqnarray}
\par
The coefficients with $\exp(i(kx_2+\omega t_2\pm 2S)/\varepsilon^2)$ and their complex conjugations are defined by:
\begin{equation}
u_{3,(j,\pm2)}=-\frac{1}{2 L[\chi_{j,\pm2}]} f u_{2,(j,\pm1)},\quad j=\pm1 \label{algE31}
\end{equation}
\par
The following equations and their complex conjugations define the term with coefficient $\exp(3(kx_2+\omega t_2)/\varepsilon^2)$ and $\exp(-3(kx_2+\omega t_2)/\varepsilon^2)$ accordingly.
\begin{equation}
u_{3,(3,0)}=- {1\over L[\chi_{3,0}]}\gamma u_1^3.\label{algE31}
\end{equation}
\par
As a result one 
obtains $U_3$ where
$$
\Omega_3=\{(1,\pm m);(-1,\pm m), \,m=1,2;(\pm3,0)\}.
$$
\par
The third correction $U_3$ has the singularity of the second order with respect to $L[(k x_2+\omega t)\pm S]$ on the curves (\ref{singCurvePlus1}) and (\ref{singCurveMinus1}), and the singularity of the first order with respect to $L[(k x_2+\omega t)\pm 2S]$ on the curves:
\begin{eqnarray}
L[(k x_2+\omega t)+2S]=0,
\quad L[(k x_2+\omega t)-2S]=0.
\label{singCurvePlus2Minus2}
\end{eqnarray}
\par 
Let us consider equation (\ref{n-thPreresonantEq}) for $n>3$. The right-hand side of (\ref{n-thPreresonantEq}) contains the phases $\chi_{j,m}$, where $(j,m)\in\Omega_{n}$ and 
\begin{eqnarray*}
\Omega_n=\Omega_{n-1}\cup
\{(j_{n-1},m_{n-1}\pm1),
(j_{n-1},m_{n-1})\in\Omega_{n-1}\}
\cup
\\
\{(j_{l_1}+j_{l_2}+j_{l_3},
m_{l_1}+m_{l_2}+m_{l_3}),
l_1+l_2+l_3=n,\,(j_{l_q},m_{l_q})\in
\Omega_{l_q},l_q\in{\mathbf N}\}.
\end{eqnarray*}
\par
The amplitudes are defined by the following formula as $(j,m)\not\equiv(\pm1,0)$:
\begin{eqnarray}
u_{n,(j,m)}={1\over L[\chi_{j,m}]}\bigg(
-i\partial_{t_2}\chi_{j,m}u_{n-1,(j,m)}+ 
\nonumber \\
\bigg[2i\pt_{t_2}\chi_{j,m} 
\pt_{t_1}u_{n-1,(j,m)}-
2i\pt_{x_2}\chi
\pt_{x_1}u_{n-1,(j,m)}\bigg]-
\nonumber \\
\frac{1}{2} f u_{n-1,(j,m-1)}-
\frac{1}{2} f u_{n-1,(j,m+1)}-
\nonumber \\
\gamma\sum_{\begin{array}{c}
q_1+q_2+q_3=n
\\(j_p,m_p)\in\Omega_{q_p}
\\j=j_1+j_2+j_3\\
m=m_1+m_2+m_3\end{array}} 
u_{q_1,(j_1,m_1)}u_{q_2,(j_2,m_2)}
u_{q_3,(j_3,m_3)}\bigg).
\label{n-th-j-m}
\end{eqnarray}
If $(j,m)\equiv(\pm1,0)$ then the functions $u_{n-1,(\pm1,0)}\equiv u_{n-1,(\pm1,0)}(t_2,\zeta)$. Function 
$u_{n,(\pm1,0)}(t_2,\zeta)$ stay indefinite and $u_{n-2,(\pm1,0)}$ is a solution of the equation:
\begin{eqnarray}
2i\omega \partial_{t_2}u_{n-2,(\pm1,0)}
+\partial_\zeta^2u_{n-2,(\pm1,0)}+ 
\nonumber \\
\gamma\sum_{\begin{array}{c}
q_1+q_2+q_3=n
\\(j_p,m_p)\in\Omega_{q_p}
\\j_1+j_2+j_3=\pm1\\
m_1+m_2+m_3=0\end{array}}
u_{q_1,(j_1,m_1)}u_{q_2,(j_2,m_2)}
u_{q_3,(j_3,m_3)}=0.
\label{n-th-1-n-2}
\end{eqnarray}
\par
The term $U_n$ has the singularity on the curve $L[\chi_{j,m}]$ of the order $p=n-(|j|+|m|)+1$. It means that the series for the solution loses the asymptotic property in a neighborhood of this curve. 
\par
The series for the solution has an asymptotic property if 
$$
{\varepsilon^{n+1}U_{n+1}\over \varepsilon^{n}U_{n}}\ll 1.
$$
It yields the domain of validity for the constructed asymptotic solution:
$$
\varepsilon L[\chi_{j,m}]\ll1.
$$
The Theorem \ref{ex1Theorem} is proved.

\section{Internal expansion}
\subsection{Expansion near primary resonant curve}
\label{secExpPrimaryResonantCurve}
\par
The primary singularity is situated at the curves $L[\chi_{1,\pm1}]=0$. It is easy to see that  the $n$-th coefficients of expansion (\ref{ext1}) has the singularity of the order $n-1$ on these  curves. 
\par
A typical local resonance generates new harmonics with 
$$
k_1=k\pm \pt_{x_2}S|_{L[\chi_{1,\pm1}]=0},\quad 
\hbox{as}\quad |k_1|\not= k
$$
and 
$$
\omega_1=\omega\pm \pt_{t_2}S|_{L[\chi_{1,\pm1}=0}.
$$
That case may be studied by the same approach as \cite{kg1}.
\par
Here we consider a special case of the local resonance when 
$$
k-\pt_{x_2}S=-k.
$$
This special case we call by the local parametric resonance. In this case a new harmonics are not generated in the leading-order term of the asymptotic expansion but the envelope function changes.
\par
In this subsection we construct the formal asymptotic solution near $L[\chi_{1,1}]=0$  and obtain an connection formulas for the nearest-neighbor solution (\ref{ext1}). 
\par
\begin{theorem}\label{inTheorem}
In a neighbourhood of the 
parametric resonant curve the formal 
asymptotic expansion for 
the solution of (\ref{pdkg}) 
modulo $O(\ve^{N+1})$ has the form
\begin{equation}
U(x,t,\ve)= \sum_{n=1}^N
\varepsilon^n 
W_n(x_1,t_1,x_2,t_2,\varepsilon), 
\label{internalSolution}
\end{equation}
where
\begin{eqnarray*}
W_n=\sum_{j=-n+1}^{n}
w_{n,j}(x_1,t_1)
\exp(i{(2j-1)S\over2\ve^2}),
\end{eqnarray*}
Function $w_{n,j}$  is determined from equations (\ref{leadingInside}), (\ref{n-thCorrectionEqInside}) and (\ref{eqForNk-thCorrectionTerm}). This expansion 
is valid when
$$
|(\pt_{t_2}S)^2-(\pt_{x_2}S)^2-4| \ll \ve^{-1}.
$$
\end{theorem}

\subsection{Formal Construction}

In this section we prove the Theorem 
\ref{inTheorem}. 
\par
Substitute (\ref{internalSolution}) into (\ref{pdkg}). Define:
\begin{equation}
\varepsilon\lambda=      
-{1\over4}(\pt_{t_2}S)^2+
{1\over4}(\pt_{x_2}S)^2+1.
\label{definitionOfLambda}
\end{equation}
Gather the therms of same order of $\varepsilon$. In order of $\varepsilon^2$ it yields:
\begin{eqnarray}
\sum_{j=-1, j\not=0,1}^2
\bigg[-{1\over4}((2j-1)\pt_{t_2}S)^2+
{1\over4}((2j-1)\pt_{x_2}S)^2+1\bigg]\times
\nonumber
\\
w_{2,j}\exp(i{(2j-1)S\over2\ve^2})=
{1\over 2}w_{1,1}f \exp\big(i{3S\over 
2\varepsilon^2}\big)+
{1\over 2}w_{1,-1}f \exp\big(-i{3S\over 
2\varepsilon^2}\big)
\nonumber
\\
\big[(i\pt_{x_2}S\pt_{x_1}-\pt_{t_2}S\pt_{x_1})w_{1,1}
+\lambda w_{1,1}+
\nonumber
\\
{f\over2} \overline{w_{1,1}}\big]
\exp(i{S\over2\ve^2})+
\nonumber
\\
\big[(i\pt_{x_2}S\pt_{x_1}-\pt_{t_2}S\pt_{x_1})w_{1,-1}
+\lambda w_{1,-1}+
\nonumber
\\
{f\over2} \overline{w_{1,-1}}\big]
\exp(-i{S\over2\ve^2}),
\nonumber
\end{eqnarray}
This formula contains linear independent exponents with respect to fast variables. Collect the coefficients of such exponents. As a result one obtains equations for $w_{1,\pm1}$ 
\begin{equation}
i(\pt_{x_2}S\pt_{x_1}-
\pt_{t_2}S\pt_{t_1})w_{1,\pm1}
+\lambda w_{1,\pm1}+
{f\over2} \overline{w_{1,\pm1}}=0,
\label{leadingInside}
\end{equation}
and a formula for $w_{2,\pm3}$:
$$
w_{2,\pm3}={1\over 4}w_{1,\pm1}f.
$$
\par
In order of $\varepsilon^n$ we obtain the following equation: 
\begin{eqnarray}
\sum_{\begin{array}{c}j=-n+1\\
j\not=0,1\end{array}}^n
\bigg[-{1\over4}((2j-1)\pt_{t_2}S)^2+
\nonumber
\\
{1\over4}((2j-1)\pt_{x_2}S)^2+1\bigg]\times
w_{n,j}\exp(i{(2j-1)S\over2\ve^2})=
\nonumber
\\
\big[(i\pt_{x_2}S\pt_{x_1}-\pt_{t_2}S\pt_{x_1})w_{n-1,1}
+\lambda w_{n-1,1}+ 
\nonumber
\\
\frac{1}{2}f \overline{w_{n-1,1}}\big]
\exp(i{S\over2\ve^2})+
\nonumber
\\
\big[(i\pt_{x_2}S\pt_{x_1}-\pt_{t_2}S\pt_{x_1})w_{n-1,-1}
+\lambda w_{n-1,-1}+ 
\nonumber
\\
\frac{1}{2}f \overline{w_{n-1,-1}}\big]
\exp(-i{S\over2\ve^2})+
\nonumber
\\
-\sum_{\begin{array}{c}j=-n+1\\
j\not=0,1\end{array}}^n 
\bigg(\pt_{t_2}^2w_{n-4,2j-1}+ 
\nonumber
\\
i(2j-1)\pt_{t_2}S\pt_{t_2}w_{n-2,2j-1}+
\nonumber
\\
\pt_{t_1}^2w_{n-2,2j-1}-
\pt_{x_1}^2w_{n-2,2j-1}+
\nonumber
\\
\frac{f}{2}w_{n-1,2j-3}+
\frac{f}{2}w_{n-1,2j+1}+
\nonumber
\\
\gamma
\sum_{\begin{array}{c}p+q+r=n\\
j_p+j_q+j_r=2j-1\end{array}}w_{p,j_p}w_{q,j_q}w_{r,j_r}
\bigg)\exp(i{(2j-1)S\over2\ve^2}).
\label{firstCorrectionEq}
\end{eqnarray}
\par
Collect coefficients with linear independent exponents with respect to fast variables. It yields:
\begin{equation}
i\pt_{t_2}S\pt_{t_1}w_{n,1} - 
i\pt_{x_2}S\pt_{x_1}w_{n,1} 
+ \lambda w_{n,1} + 
\frac{1}{2}f \overline{w_{n,1}} =
F_{n,1} 
\label{n-thCorrectionEqInside}
\end{equation}
where
\begin{eqnarray}
{F}_{n,1}=
 -i\pt_{t_2}S\pt_{t_2}{w}_{n-1,1} +
i\pt_{x_2}S\pt_{x_2}{w}_{n-1,1}+
\nonumber\\
{1\over4}(\pt_{t_2}S)^2{w}_{n-1,1} -
{1\over4}(\pt_{x_2}S)^2{w}_{n-1,1}-
\nonumber\\
-\pt_{t_1}^2{w}_{n-1,1} +
\pt_{x_1}^2{W}_{n-1,1}-
2\pt_{t_2}\pt_{t_1}{w}_{n-2,1}+
2\pt_{x_2}\pt_{x_1}{w}_{n-2,1}-
\nonumber \\
- \pt_{t_2}^2{w}_{n-3,1} + \pt_{x_2}
^2{w}_{n-3,1}-
\nonumber\\
\g\sum_{\begin{array}{c} n_1+n_2+
n_3=n+1 , \\
 k_1 + k_2 +k_3 =1\\ k_j\in 
\Omega_{n_j},\,j=1,2,3 \end{array}}
{w}_{n_1,k_1}
{w}_{n_2,k_2}
{w}_{n_3,k_3}.
\label{rightSideOfEqForN1-thCorrectionTerm}
\end{eqnarray}
\par
The term ${w}_{n,j}$, $j\not=1$ is 
determined by algebraic equation
\begin{eqnarray}
{w}_{n,j}= \frac{\gamma}{(2j-1)^2 -1}
\left(
-2i\pt_{t_2}S\pt_{t_2}{w}_{n-2,j} +
2i\pt_{x_2}S\pt_{x_2}{w}_{n-2,j}+
\right.
\nonumber\\
(\pt_{t_2}S)^2{w}_{n-2,j} -
(\pt_{x_2}S)^2{w}_{n-2,j}-
\pt_{t_1}^2{w}_{n-2,j} +
\pt_{x_1}^2{w}_{n-2,j}-
\nonumber\\
2\pt_{t_2}\pt_{t_1}{w}_{n-3,j}+
2\pt_{x_2}\pt_{x_1}{w}_{n-3,j}-
- \pt_{t_2}^2{w}_{n-4,j} + 
\pt_{x_2}^2{w}_{n-4,j}-
\nonumber\\
\sum_{\begin{array}{c} n_1+n_2+n_3
=n+1 , \\
k_1 + k_2 +k_3 =j\\ k_j\in 
\Omega_{n_j},\,j=1,2,3
\end{array}}
 {w}_{n_1,k_1}
{w}_{n_2,k_2} {w}_{n_3,k_3}\Bigg).
\label{eqForNk-thCorrectionTerm}
\end{eqnarray}

\subsubsection{Characteristic variables}

The function ${w}_{n,1}$ satisfies equation (\ref{leadingInside}). The solution is constructed by the method of characteristics. Define the characteristic variables $\s,\xi$. We choose a point $(x^0_1,t^0_1)$ such that $\pt_{x_2}l|_{(x^0_1,t^0_1)}\not=0$ as an origin and denote by $\s$ the variable along the characteristics for equation (\ref{leadingInside}). We suppose  $\s=0$ on the curve $\l=0$. The variable $\xi$ mensurates the distance along the curve $\l=0$ from the point $(x^0_1,t^0_1)$. This point $(x^0_1,t^0_1)$ corresponds to $\xi=0$. The positive direction for parameter $\xi$ coincides with the positive direction of $x_2$  in the neighborhood of $(x^0_1,t^0_1)$. 
\par
The characteristic equations for 
(\ref{leadingInside}) have a
form
\bb
{dt_1\over d\s}=\pt_{t_2}
S(\ve x_1,\ve t_1),\quad {dx_1\over
d\s}=-\pt_{x_2}S(\ve x_1,\ve t_1).
\label{eqForCharacteristics}
\ee
The initial conditions for the equations 
are
\bb
x_1|_{\s=0}=x^0_1,\quad
t_1|_{\s=0}=t^0_1.
\label{initialConditionsForCharacteristics}
\ee
\begin{lemma}
\label{lemmaAboutSolvabilityForCharateristicEq}
The Cauchy problem for characteristics 
has a solutions when
 $|\s|<c_1\ve^{-1},\quad 
c_1=const>0$.
\end{lemma}
{\bf Proof.} The Cauchy problem 
(\ref{eqForCharacteristics}),
(\ref{initialConditionsForCharacteristics}) 
is equivalent  to the
system of the integral equations
\bb
t_1=t^0_1+\int_0^\s  
\pt_{t_2}S(\ve x_1,\ve t_1)d\z,
\quad
x_1=x^0_1-\int_0^\s  
\pt_{x_2}S(\ve x_1,\ve t_1)d\z.
\label{integralEqForCharacteristics}
\ee
Substituting  
$\tilde t_2=(t_1-t^0_1)\ve,\,\,\tilde
x_2=(x_1-x^0_1)\ve$, we obtain
$$
\tilde t_2=\int_0^{\ve \s}  
\pt_{t_2}S(\tilde x_2-\ve x^0_1,
\tilde
t_2-\ve t^0_1)d\z,\quad \tilde x_2=-
\int_0^{\ve\s}
 \pt_{x_2}S(\tilde x_2-\ve x^0_1,
\tilde t_2-\ve t^0_1)d\z.
$$
The integrands are smooth and 
bounded functions on the plane 
$x_2,t_2$. There exists the constant 
$c_1=\const>0$ such that the integral 
operator is a contraction operator when 
$\ve|\s|<c_1$. Lemma  
\ref{lemmaAboutSolvabilityForCharateristicEq} 
is proved. 
\par
It is convenient to use the following 
asymptotic formulas for the change
of variables  $(x_1,t_1)\to(\s,\xi)$.
\par
\begin{lemma}
\label{lemmaAboutAsymptoticsForCharacteristics}
In the domain $|\s|\ll\ve^{-1}$ the 
asymptotics  as $\ve\to0$ of the
solutions for Cauchy problem 
(\ref{eqForCharacteristics}),
(\ref{initialConditionsForCharacteristics})  have the form
\begin{eqnarray}
x_1(\s,\xi,\ve)-x^0_1(\xi)=-
\s \pt_{x_2}S+\sum_{n=1}^N
\ve^n\s^{n+1} g_n(\ve x_1,\ve t_1)+
O(\ve^{N+1}\s^{N+2}),\qquad
\label{asymptoticsOf-x1}\\
t_1(\s,\xi,\ve)-t^0_1(\xi)=\s 
\pt_{t_2}S+\sum_{n=1}^N
\ve^n\s^{n+1} h_n(\ve x_1,\ve t_1)+
O(\ve^{N+1}\s^{N+2}),\qquad
\label{asymptoticsOf-t1}
\end{eqnarray}
where
$$
g_n=-{d^n \over d \s^n}(\pt_{x_2}S)
\bigg|_{\s=0},\quad  h_n={d^n
\over d \s^n}(\pt_{t_2}S)\bigg|_{\s=0}.
$$
\end{lemma}
\par
The lemma proves by integration by 
parts of equations
(\ref{integralEqForCharacteristics}). 
\par
The next proposition gives us the 
asymptotic formula which relates the 
variables $\s$ and $\l$ as $\s,\l\to \pm
\infty$.

\begin{lemma}
\label{lemma_lambda_and_sigma}
Let be $\s\ll \ve^{-1}$, then:
$$
\l=\varphi(\xi)\s+O(\ve\s^2),\quad 
\s\to \infty,\quad \varphi(\xi)=
{d\l\over
d\s}\bigg|_{\s=0}.
$$
\end{lemma}
\par
{\bf Proof.} From formula 
(\ref{definitionOfLambda}) we 
obtain
the representation in  the form
$$
\l=\sum_{j=1}^\infty \l_j(x_1,t_1,\ve)
\s^j\ve^{j-1},
$$
where
$$
\l_j(x_1,t_1,\ve)={1\over j!}{d^j\over
d\s^j}\l(x_1,t_1,\ve)|_{\s=0}.
$$
\par
It yields
$$
\l={d\l\over d\s}\big|_{\s=0} \s +
O\big(\ve\s^2 {d^2 \l\over
d\s^2}\big).
$$
Let be
$$
\left|{d^2 l\over d\s^2}\right|\ge
\const, \, \, \xi\in R.
$$
\par
The function $d\l/ d\s$ is not equal to  
zero
$$
{d\l\over
d\s}={1\over2}\bigg(-\pt_{x_2}\l
\pt_{x_2}S+\pt_{t_2}\l\pt_{t_2}S
\bigg)\not=0.
$$
Let us suppose $d\l/ d\s>0$. It yields
$$
\l=\varphi(\xi)\s+O(\ve\s^2),\quad 
\quad \varphi(\xi)={d\l\over
d\s}\bigg|_{\s=0}
$$
The lemma is proved.

\subsubsection{Ordinary differential 
equations for local parametric resonance}
\par
The equation (\ref{leadingInside}) may be considered as an ordinary differential equation along the characteristic.
\begin{equation}
i{dw_{1,1}\over d\sigma}+
\lambda(\sigma,\xi,\varepsilon) w_{1,1}+
{f\over 2}\overline{w}_{1,1}=0.
\label{eqw11}
\end{equation}
Let us expand $\lambda$ as a series with respect to $\varepsilon$ and consider (\ref{eqw11}) as a perturbation of:
$$
i{dw_{1,1}\over d\sigma}+
\varphi(\xi)\sigma w_{1,1}+
{f\over 2}\overline{w}_{1,1}=0.
$$
Let us change variable:
$\kappa=\alpha \sigma$, 
$F=f/(2\alpha)$ where $\alpha^2(\xi)=
\varphi(\xi)$, 
then we obtain:
\begin{equation}
i{d w_{1,1}\over d\kappa}+\kappa 
w_{1,1}+F\overline{w_{1,1}}=0.
\label{PLRE}
\end{equation} 
\par
The same equation for $w_{n,1}$ has the following form:
\begin{equation}
i{d w_{n,1}\over d\kappa}+\kappa 
w_{n,1}+F\overline{w_{n,1}}=
{1\over\alpha}F_{n,1}+
{1\over\alpha}\sum_{j+l=n-1}
\lambda_{l+1}\bigg({\kappa\over\alpha}
\bigg)^{l+1}w_{j,1}. 
\label{PLRE-n}
\end{equation}
\par 
A solution for the equation (\ref{PLRE}) 
has obtained in \cite{BuslaevDmitrieva}:
\begin{eqnarray}
V(\kappa;C)={1\over2}
(-1-i)^{1- i{F^2\over2}}\, F\, 
\overline{C} D_{z}(e^{i\pi\over4}
\sqrt{2} \kappa) + 
\nonumber\\
\bigg((1-i)^{i{F^2\over2}}C+
2(-1-i)^{-1-i{F^2\over2}}
e^{{F^2\pi\over2}}{\sqrt{2\pi}\over 
F\Gamma(-i{F^2\over2})}\overline{C}
\bigg)D_{-z-1}(e^{{5\pi\over4}}
\sqrt{2} \kappa),
\nonumber
\end{eqnarray}
where $D_z(y)$ is a parabolic cylinder 
function, $z=i{F^2\over2}-1$ and $C$ 
is an arbitrary complex constant.  
\par
Let us define 
\begin{eqnarray}
w_{1,1}=V(\kappa,C).
\end{eqnarray}
\par 
Let us consider the nonhomogeneous equation:
\begin{equation}
i{d U\over d\kappa}+\kappa U+
F\overline{U}=G(\kappa).
\label{nonhomogenPLRE}
\end{equation}
To solve this equation we use two linear independent solutions of homogeneous equation. These solutions are:
$$
V_1=\pt_{\alpha}V(\kappa,C),\quad 
V_2=\pt_{\beta}V(\kappa,C),\quad 
C=\alpha+i\beta.
$$
Let us define the Wronskian of these 
solutions as:
$$
W=V_1\overline{V_2}-V_2
\overline{V_1}.
$$
It easy to see that $W=\const\not=0$.
\par
The solution of (\ref{nonhomogenPLRE}) 
has a form:
\begin{eqnarray}
U=V_1\int^\kappa (G \overline{V_2}-
V_2 \overline{G})(\chi) {d\chi\over W}
+
\nonumber\\
V_2\int^\kappa (G \overline{V_1}-V_1 
\overline{G})(\chi) {d\chi\over W}+ 
c_1 V_1+c_2 V_2,
\label{solutionsOfNonhomogenousEquation}
\end{eqnarray}
where $c_1$ and $c_2$ are real 
constants which are parameters of the 
solution.
\par
The general formula 
(\ref{solutionsOfNonhomogenousEquation}) 
allows to us to solve the equations for 
any high-order term. 

\subsection{Asymptotics as $\l\to\infty$ 
and domain of validity of
the internal expansion}
\label{ProofOfTheorem2Step3}
\par
The domain of validity of the internal 
expansion is determined by the 
asymptotic behaviour of higher-order 
terms. In this section we show, that 
the $n-$th order term of the 
asymptotic solution grows as 
$\l^{n-1}$ when $\l\to\infty$. This 
growth of higher-order terms  allows 
us to determine the domain of validity 
for internal asymptotic expansion (\ref{internalSolution}) as 
$\l\to\infty$.

\subsubsection{Asymptotic behaviour of leading-order term}
\par
The leading-order term has the following behaviour as $\kappa\to-\infty$:
\begin{eqnarray}
w_{1,1}=C_{1,1} e^{i\big({\kappa^2\over2}
-{a^2\over2}\ln(-\kappa)\big)}(1+
O(\kappa^{-1})),\quad 
C_{1,1}\in{\mathbf C}
\label{leftAs}
\end{eqnarray}
and 
\begin{eqnarray}
w_{1,1}=\bigg(e^{{a^2\pi\over2}} C_{1,1}+
{(1+i)e^{{a^2\pi\over4}}
e^{i{a^2\over4}\ln(2)}a\sqrt{\pi}
\over \Gamma(1-i{a^2\over2})} 
\overline{C_{1,1}}\bigg)
e^{i\big({\kappa^2\over2}-
{a^2\over2}\ln(\kappa)\big)}
(1+O(\kappa^{-1}))
\label{rightAs}
\end{eqnarray}
as $\kappa\to\infty$.
\par
This formulas were obtained using the 
asymptotic behavior of solution (\ref{PLRE}). 
\subsubsection{Asymptotic behavior of 
higher-order terms}
\par
The $n$-th order term has the following 
asymptotic formula:
$$
w_{1,n}=O(\kappa^{n-2}).
$$
Then the internal expansion is  
valid as $|\kappa|\ll\varepsilon^{-1}$, 
or the same as 
$\varepsilon|\lambda|\ll1$.

\section{Matching}
\par
The domains of validity for the internal and external asymptotic expansions are intersected. It allows us to match these expansions and as a result to obtain an uniform asymptotic expansion that is valid in both domains.
\par
To match the expansions one should reexpand the external asymptotic expansion in the terms of internal variables $\xi$ and $\kappa$ and equate the terms with the same order of $\varepsilon$. It yields the indefinite functions in the internal asymptotic expansion. Here we make this step for the main terms of the internal and external asymptotic expansions before the local resonant layer:
\begin{eqnarray}
\varepsilon\bigg(w_{1,-1}(\xi,\kappa)\exp(iS/\varepsilon^2)+c.c.\bigg)\sim
\nonumber\\
\varepsilon\bigg(\Psi(t_2,\zeta)\exp(-i{f^2\over 4\omega}G(x_2,t_2))\exp(i(k x_2+\omega t_2)/\varepsilon^2)+c.c.\bigg),
\label{matchingBefore}
\end{eqnarray}
as $1\ll-\kappa\ll\varepsilon^{-1}$, $(t_2,x_2)\in L[\chi_{1,1}]=0$ and $\zeta=\omega x_1^0(\xi)+k t_1^0(\xi)$.
This formula define the function $w_{1,-1}(\xi,\kappa)$ as $\kappa\to-\infty$.
\par
The same formula for the matching is valid after the resonant layer:
\begin{eqnarray}
\varepsilon\bigg(\Psi(t_2,\zeta)\exp(-i{f^2\over 4\omega}G(x_2,t_2))\exp(i(k x_2+\omega t_2)/\varepsilon^2) 
+c.c.\bigg)
\sim
\nonumber\\
\varepsilon\bigg(w_{1,-1}(\xi,\kappa)\exp(iS/\varepsilon^2)+c.c.\bigg).
\label{matchingAfter}
\end{eqnarray}
as $1\ll\kappa\ll\varepsilon^{-1}$, $(t_2,x_2)\in L[\chi_{1,1}]=0$ and $\zeta=\omega x_1^0(\xi)+k t_1^0(\xi)$.
This formula defines the function $\Psi(t_2,\zeta)$ coming out of the resonant layer. 
\par
The matching gives a jump for the $\Psi$ at the resonant line:
\begin{eqnarray}
\Psi(t_2,\zeta)|_{L[\chi_{1,1}]+0}=\bigg(e^{{f^2\pi\over8}} \Psi(t_2,\zeta)|_{L[\chi_{1,1}]-0}+
\nonumber\\
{(1+i)e^{{f^2\pi\over16}}
e^{i{f^2\over16}\ln(2)}f\sqrt{\pi}
\over 2\Gamma(1-i{f^2\over8})} 
\overline{\Psi(t_2,\zeta)}|_{L[\chi_{1,1}]-0}\bigg)
\end{eqnarray}

\subsection{Post-resonant expansion}
\par
The asymptotic structure of the post-resonant solution has the similar structire as the pre-resonant asymptotic solution. The main term of the solution is:
$$
U(x,t,\ve) \sim \ve v_1(x_1,t_1,t_2)\exp\{i(kx+\omega t) \}
$$
The amplitude $v_1=\Psi  \exp\left\{i\displaystyle\frac{f^2}{4\omega}G(x_2,t_2)\right\}$ and $\Psi$ is determined by nonlinear Schr\"odinger equation also  and initial datum on the curve $L[\chi_{1,1}]=0$
$$
\Psi(t_2,\zeta)|_{L[\chi_{1,1}]+0}=e^{{f^2\pi\over8}} \Psi(t_2,\zeta)|_{L[\chi_{1,1}]-0}+
{(1+i)e^{{f^2\pi\over16}}
e^{i{f^2\over16}\ln(2)}f\sqrt{\pi}
\over 2\Gamma(1-i{f^2\over8})} 
\overline{\Psi(t_2,\zeta)}|_{L[\chi_{1,1}]-0}.
$$

\end{document}